# Investigation of PVC plastisol tissue-mimicking phantoms for MR- and ultrasound-elastography

**Simon Chatelin**[1*], **Elodie Breton**[1], **Ajeethan Arulrajah**[1,2], **Céline Giraudeau**[2], **Benoit Wach**[1], **Laurence Meylheuc**[1,3] **and Jonathan Vappou**[1]

[1]ICube, University of Strasbourg, CNRS UMR 7357, IHU-Strasbourg, Strasbourg, France

[2]IHU, Institut Hospitalo-Universitaire, Institute for Image Guided Surgery, Strasbourg, France

[3]INSA, Institut National de Sciences Appliquées, Strasbourg, France

**\* Correspondence:**
Simon Chatelin
schatelin@unistra.fr

**Abstract**

*Objective:* Realistic tissue-mimicking phantoms are essential for the development, the investigation and the calibration of medical imaging techniques and protocols. Because it requires taking both mechanical and imaging properties into account, the development of robust, calibrated phantoms is a major challenge in elastography. Soft polyvinyl chloride gels in a liquid plasticizer (plastisol or PVCP) have been proposed as soft tissue-mimicking phantoms (TMP) for elasticity imaging. PVCP phantoms are relatively low-cost and can be easily stored over long time periods without any specific requirements. In this work, the preparation of a PVCP gel phantom for both MR and ultrasound-elastography is proposed and its acoustic, NMR and mechanical properties are studied.

*Material and methods:*

The acoustic and magnetic resonance imaging properties of PVCP are measured for different mass ratios between ultrasound speckle particles and PVCP solution, and between resin and plasticizer. The linear mechanical properties of plastisol samples are then investigated over time using not only indentation tests, but also MR and ultrasound-elastography clinical protocols. These properties are compared to typical values reported for biological soft tissues and to the values found in the literature for PVCP gels.

*Results and conclusions:* After a period of two weeks, the mechanical properties of the plastisol samples measured with indentation testing are stable for at least the following 4 weeks (end of follow-up period 43 days after gelation-fusion). Neither the mechanical nor the NMR properties of plastisol gels were found to be affected by the addition of cellulose as acoustic speckle. Mechanical properties of the proposed gels were successfully characterized by clinical, commercially-available MR Elastography and sonoelastography protocols. PVCP with a mass ratio of ultrasound speckle particles of 0.6% to 0.8% and a mass ratio between resin and plasticizer between 50 and 70% appears as a good TMP candidate that can be used with both MR and ultrasound-based elastography methods.

**Investigation of PVC plastisol tissue-mimicking phantoms for MR- and ultrasound-elastography**

## 1 Introduction

Over the past three decades, different methods have been developed for tissue elasticity measurement using medical imaging. All elastography approaches rely on the encoding of tissue displacement as a result of a force field that can be either external or internal, static or dynamic (Vappou, 2012). Displacements can be encoded by using medical imaging, such as ultrasound imaging, Magnetic Resonance Imaging (MRI) or optical imaging. Tissue-mimicking phantoms (TMP) are of primary importance during the development, validation and calibration processes in elasticity imaging (Oudry *et al.*, 2014; Cao *et al.*, 2017) and during operator training phases. Research-based and commercially-available TMP are today proposed for elastography, and are mostly dedicated to one specific elastography modality (Cournane, Fagan and Browne, 2012).

Ideally, elastography-dedicated TMP are expected to offer the following features: (1) Their mechanical properties (such as elasticity, viscosity, anisotropy, porosity or hyperelasticity) must be well controlled and must lie within typical values of soft tissues they are supposed to mimic; (2) They should offer particular ease of use in terms of storage conditions and durability; (3) They must be compatible with the medical imaging modality for which they have been developed. Several studies have proposed elastography TMP that display particular mechanical properties beyond linear elasticity, such as viscosity (Madsen *et al.*, 2003; Hadj Henni *et al.*, 2011; Nguyen *et al.*, 2014), poroelasticity (Chaudhry *et al.*, 2016), anisotropy (Namani *et al.*, 2009; Qin *et al.*, 2013; Aristizabal *et al.*, 2014; Chatelin *et al.*, 2014; Schmidt *et al.*, 2016), hyperelasticity (Erkamp *et al.*, 2004; Gennisson *et al.*, 2007; Pavan *et al.*, 2010), heterogeneity (Gao *et al.*, 1995; Bishop *et al.*, 2000; Plewes *et al.*, 2000; Madsen *et al.*, 2005; Green, Bilston and Sinkus, 2008; Mariappan *et al.*, 2009) and TMP including dynamic flow pulsations (Kolipaka *et al.*, 2009). Inclusion of anisotropy (Chatelin *et al.*, 2014) or porosity could be suggested through the addition of fibrin fibers or through the use of 3D-printing, as proposed by several authors for TMP (Wan *et al.*, 2014; Guidetti *et al.*, 2019). Important features that need to be accounted for are the preservation process, the stability over time and the inhalation toxicity during manufacturing. Depending on the imaging method, elastography TMP are mainly composed of a matrix, solvents and other additives (Cao *et al.*, 2017). Studies comparing elastography measurements obtained with different imaging modalities have illustrated the need for multi-modality elastography TMP (Oudry, Chen, *et al.*, 2009; Brinker *et al.*, 2018).

Synthetic phantoms have very interesting properties and are widely used as TMP. The most commonly used synthetic TMP arestyrene-ethylene/butylene-styrene (SEBS) (Oudry, Bastard, *et al.*, 2009; Oudry *et al.*, 2014) and silicone-based materials (more specifically Ecoflex gels) (Hadj Henni *et al.*, 2011; Piazza *et al.*, 2012; Yasar, Royston and Magin, 2013; Liu, Yasar and Royston, 2015). Soft PolyVinyl Chloride (PVC) suspensions in a liquid plasticizer (plastisol or PVCP) have also been proposed as tissue-mimicking candidates for elasticity imaging (Bishop *et al.*, 2000; Hadj Henni *et al.*, 2011; Leclerc, Charleux, *et al.*, 2012; Leclerc, Debernard, *et al.*, 2012; Chartier *et al.*, 2014; Montagnon *et al.*, 2014; Chakouch *et al.*, 2015; Lefebvre *et al.*, 2016; Arunachalam *et al.*, 2017). These TMP are good candidates for mimicking different stages of liver fibrosis, by varying their mass ratio between resin and plasticizer $M_{Res/Plast}$ from 40% to 60% (Chakouch *et al.*, 2015). Soft PVCP has been proposed in robotics as TMP for needle insertion (DiMaio and Salcudean, 2003; Hungr *et al.*, 2012), and also for Magnetic Resonance Elastography (MRE) (Bishop *et al.*, 2000, 2001; Samani, Bishop and Plewes, 2001; Leclerc, Charleux, *et al.*, 2012; Leclerc, Debernard, *et al.*, 2012; Chakouch *et al.*, 2015; Lefebvre *et al.*, 2016; Brinker *et al.*, 2018), sonoelastography (Montagnon *et al.*, 2014; Chakouch *et al.*, 2015) and biomedical photoacoustic (Spirou *et al.*, 2005; Bohndiek *et al.*, 2013; Fonseca *et al.*, 2016; Vogt *et al.*, 2016; Jia *et al.*, 2017; Dantuma, van Dommelen and Manohar, 2019). PVCP has been previously compared to silicone-based Ecoflex (Brinker *et al.*, 2018) and has been shown to be a very interesting

**Investigation of PVC plastisol tissue-mimicking phantoms for MR- and ultrasound-elastography**

candidate for MRE in terms of MRI and dynamic viscoelastic properties (Li *et al.*, 2016; Arunachalam *et al.*, 2017; He *et al.*, 2019). In addition to their relatively low cost, PVCP TMP are simple to prepare, stable over time at room temperature and resistant to damage caused by typical handling procedures.

However, whether the same phantoms can be used simultaneously for sonoelastography and MRE remains poorly known. This study aims at investigating several key properties of the same PVCP phantoms for application to MRE and sonoelastography. In particular, the influence of added particles for ultrasound speckle on acoustic properties, MRI relaxation times and mechanical properties is investigated. Long term stability in mechanical properties is investigated over 43 days, and phantoms with different resin and speckle concentrations are tested using commercial MR and ultrasound elastography systems.

## 2   Material and methods

### 2.1   Description of PVCP samples

PVCP is widely used in industry, mainly textile, automobile and aeronautics. In this study, we consider its use specifically for MR and ultrasound elasticity imaging. PVCP is a combination of a PVC colloidal suspension in a liquid ester plasticizer, the role of which is to soften the final material. These gels are formed through the gelation-fusion process (Nakajima, Isner and Daniels, 1981). In the current study, after mixing the PVC resin suspension (40% to 80%) and the plasticizer (a Bis(2-ethylhexyl) adipate) (*Plastileurre Soft and Softener*, *Bricoleurre*, *Mont-Saint-Aignan*, *France*), the solution is heated. Complete gelation-fusion requires to reach a temperature of 160°C while the thermal degradation limit of PVCP is 190°C (Bakaric *et al.*, 2020), hence the curing temperature should remain between 160 and 190°C. Curing temperature can be achieved for instance through hot plate heating in an open beaker (Hungr *et al.*, 2012) or through bain-marie oil bath heating (Bohndiek *et al.*, 2013; Vogt *et al.*, 2016; Dantuma, van Dommelen and Manohar, 2019). In this study, the solution is heated in an open glass beaker by means of a microwave oven (800W) with regular stirring (total heating duration of about 8 minutes including 5 seconds stirring every 2 minutes for the first 6 minutes and then every 20 seconds for the last 2 minutes) until reaching a target temperature of 180 °C. Cellulose particles (*Sigmacell Cellulose, Cotton linters type 50, 50µm, SigmaAldrich, Saint Louis, MI, USA*) are added as ultrasound speckle particles, once the gel has cooled down to a temperature of 80°C in order to avoid thermal degradation of cellulose. The gel is stirred with mass concentrations of 0%, 0.6%, 0.8% and 1% of cellulose particles. The solution is then degassed in a vacuum bell for 5 minutes and, poured into cylindrical molds. This process was repeated in order to make cylindrical samples of two different sizes: diameter 60 mm and height 30 mm for investigation of the imaging properties (ultrasound attenuation and MR relaxation times) and diameter 98 mm and height 75 to 90 mm for investigation of the mechanical properties (indentation, MRE and sonoelastography). The mass ratio between resin and plasticizer (mass fraction noted $M_{Res/Plast}$) as well as mass ratio between cellulose and PVCP solution (mass fraction noted $M_{Cell/Plast}$) can influence the mechanical and imaging properties of the samples, as hereafter investigated. Each sample is tested at the earliest 1 day after preparation (day #1).

### 2.2   Influence of acoustic speckle on the imaging properties of PVCP

#### 2.2.1 Influence on the ultrasound attenuation

The acoustic attenuation of PVCP samples is investigated using the method proposed in (He and Zheng, 2001). The reader is referred to Appendix A for more details.

**Investigation of PVC plastisol tissue-mimicking phantoms for MR- and ultrasound-elastography**

In the present study, we consider the acoustic attenuation properties of PVCP at a single nominal frequency of 1 MHz. The density of water, the speed of sound in water, the density of the specimen (regardless of mass ratio between resin and plasticizer $M_{Res/Plast}$, in accordance with (Chakouch, 2016), who used PVCP from the same manufacturer) and the speed of sound in the specimen are assumed to have values of 1 000 kg.m$^3$, 1 500 m.s$^{-1}$, 1 000 kg.m$^3$ and 1 400 m.s$^{-1}$, respectively (values from (Hungr et al., 2012), who used PVCP from the same manufacturer, and in accordance with (Madsen et al., 2003; Spirou et al., 2005; Fonseca et al., 2016; Bakaric et al., 2020)). The values for the density and speed of sound are close to those of most biological soft tissues (1050 kg.m$^{-1}$ and 1578 m.s$^{-1}$ for liver, 1041 kg.m$^{-1}$ and 1550 to 1630 m.s$^{-1}$ for muscle, 1035 kg.m$^{-1}$ and 1562 m.s$^{-1}$ for brain, 928 kg.m$^{-1}$ and 1430 to 1450 m.s$^{-1}$ for fat, (Duck, 1990; Culjat et al., 2010)).

The acoustic test bench is composed of a single element ceramic focused transducer (23 mm diameter, nominal frequency of 1 MHz, PA-765, *Precision Acoustics Ltd, Dorchester, UK*) driven by a digital wave generator (33210A, *Agilent Technologies Inc., Santa Clara, CA, USA*), a 0.2 mm needle hydrophone (SN2319, *Precision Acoustics Ltd, Dorchester, UK*) connected in series with a preamplifier and an oscilloscope (TDS 2002B, *Tektronix Inc., Beaverton, OR, USA*). More details on both the measurement method and the experimental set-up are proposed and illustrated in the appendix A.

### 2.2.2 Influence on the nuclear magnetism relaxation times

The potential of PVCP as a TMP for MRE is assessed through the evaluation of its NMR relaxation times ($T_1$ and $T_2$). The influence of the mass ratio between resin and plasticizer $M_{Res/Plast}$ and of the cellulose (ultrasonic speckle) mass ratio $M_{Cell/Plast}$ on the NMR relaxation times is evaluated.

The acquisitions are performed in a 1.5 T MRI scanner (MAGNETOM Aera, *Siemens AG, Erlangen, Germany*) at room temperature (22°C) on samples with mass ratio between resin and plasticizer $M_{Res/Plast}$ of 40%, 50%, 60% and 70%. The additional influence of ultrasonic speckle on the NMR relaxation times is also investigated by adding 0.6%, 0.8% and 1%-concentrated cellulose. Overall, 16 samples are imaged.

The spin-lattice relaxation time $T_1$ at 1.5 T is evaluated using a turbo spin echo sequence (TE/TR 6/3 000 ms, turbo factor 7, image acquisition time 97 s) with selective inversion recuperation preparation pulse and varying inversion times (TI) of 23, 50, 100, 150, 200, 250, 300, 350, 400, 500, 600, 800, 1 000, 1 500, 2 000 and 2 970 ms. The spin-spin relaxation time $T_2$ is evaluated using a spin echo sequence (TR 2500 ms, partial Fourier 6/8, acquisition time 363 s) with varying echo times (TE) 3.5, 5, 10, 15, 20, 30, 40, 50, 60, 80, 100, 150 and 200 ms. Common relevant imaging parameters for both sequences are: matrix 192 x 192, field of view 340 mm x 340 mm, slice thickness 6 mm, bandwidth 789 Hz.px$^{-1}$.

### 2.3 Mechanical properties

Based on the observations reported in the previous sections, standard MRE and sonoelastography (Acoustic Radiation Force Imaging - ARFI) protocols used *in vivo* in clinical practice are investigated in the same TMP and compared to a reference mechanical testing approach, namely, indentation. The mechanical stability over time of the gels is reported using these three modalities at day #1 and day #43 after the beginning of the gelation-fusion process. The mass fraction of cellulose particles to PVCP solution $M_{Cell/Plast}$ is fixed to 0.6% based on previous measurements of the acoustic attenuation.

**Investigation of PVC plastisol tissue-mimicking phantoms for MR- and ultrasound-elastography**

### 2.3.1 Indentation measurements

The linear elastic behavior is characterized against mass ratio between resin and plasticizer $M_{Res/Plast}$ from 50% to 80% with 10% steps.

The acquisitions are performed using a home-made dedicated indentation set-up (20 mm-diameter hemispherical shaped indenter mounted on a 1 degree-of-freedom translation motor and a force sensor) at room temperature (22°C) on cylindrical-shaped samples (98 mm and 75 to 90 mm in diameter and height, respectively). They are indented up to 11 mm depth at a speed of 7.6 mm.s$^{-1}$ (corresponding to strain rates between 0.08 and 0.1 s$^{-1}$). For this range of deformations, the mechanical behavior of PVCP can be considered linear (Hungr *et al.*, 2012). More details on the experimental set-up are proposed and illustrated in the appendix B. The elastic modulus E is computed using the continuous stiffness measurement (Oliver and Pharr, 1992) method and the Sneddon relationship (Sneddon, 1965), as described in (Loubet *et al.*, 1984). In order to investigate the capability of PVCP to mimic biological soft tissues for shear wave elastography and to compare with the results from MRE and ARFI measurements, the shear modulus G is deduced from: $G = E/2(1 + \nu) \approx E/3$. Each measurement is repeated five times and the mean values and standard deviations over these five measurements are reported.

In order to evaluate the stability over time of mechanical properties, the indentation method is performed in the same samples over a period of 43 days (every 3 to 14 days) after the beginning of the gelatin-fusion process. The temperature is constant and equal to $T_0 \approx 22°C$ for both storage and testing conditions.

### 2.3.2 ARFI measurements

Shear wave velocity $c_s$ measurements are performed using Siemens Acuson S3000 ultrasound imager together with a linear 9-L4 ultrasound transducer probe in the virtual touch quantification (VTq) ARFI mode (*Siemens Medical Inc., Mountain View, CA, USA*). For each sample, 10 measurements are performed and averaged, distributed along two lines at depths of 15 mm and 30 mm. To allow multi-modality comparison, the shear modulus is considered as $G = \rho c_s^2$.

### 2.3.3 MRE measurements

MRE acquisitions are performed on a 1.5 T MRI scanner (*MAGNETOM Aera, Siemens AG, Erlangen, Germany*) using a standard *in vivo* MRE protocol with motion sensitizing gradient. The mechanical waves are generated using a commercial pneumatic driver system (*Resoundant®, Mayo Clinic Foundation, Rochester, MN, USA*). Scanning parameters, excitation frequency $f$ of 60.1 Hz, echo time /repetition time (TE/TR) = (14.47 ms/50 ms), flip angle 25°, slice thickness 7 mm, acquisition matrix 128 by 102, reconstruction matrix 256 by 204, resolution 1.5625 mm by 1.5625 mm, and 1 slice parallel to the vibrating plate. Assuming pure elasticity and that the stiffness is defined using the scalar shear modulus $G = \rho(\lambda f)^2$ ($\lambda$ being the wavelength), the mechanical parameters are estimated using the local frequency algorithm (Manduca *et al.*, 2001). The results are given in terms of mean shear modulus and their standard deviation in cylindrical regions of interest (diameter 80 mm) at the center of each phantom.

ARFI and MRE measurements were performed at day #1 (beginning of the gelatin-fusion process) and day #43. A total of 5 gels with mass ratio between resin and plasticizer varying from 40 % to 80% with 10% steps was tested with both MRE and ARFI sonoelastography.

## 3 RESULTS

### 3.1 Influence of acoustic speckle on the ultrasound attenuation

Depending on $M_{Res/Plast}$, the attenuation values obtained at 1 MHz vary from 0.146 to 0.381, 0.458 to 0.817, 0.665 to 1.251 and 0.827 to 1.641 dB.cm$^{-1}$ for $M_{Cell/Plast}$ of 0%, 0.6%, 0.8% and 1%, respectively. The values are reported in figure 1 (and further detailed in Appendix C). The results are systematically compared to typical values found in the literature not only for biological soft tissues (for instance 0.45 dB.cm$^{-1}$ for liver, 0.5 to 1.5 dB.cm$^{-1}$ for muscle, 0.58 dB.cm$^{-1}$ for brain, 0.6 to 0.8 dB.cm$^{-1}$ for fat at 1 MHz (Duck, 1990)) but also for similar PVCP TMP, most often over a wider frequency range (Spirou *et al.*, 2005; Maggi *et al.*, 2013; De Carvalho *et al.*, 2016; Fonseca *et al.*, 2016; Vogt *et al.*, 2016; Dantuma, van Dommelen and Manohar, 2019; Bakaric *et al.*, 2020). With the exception of the values obtained for a mass ratio of resin to plasticizer $M_{Res/Plast}$ of 50%, the acoustic attenuation values increase with $M_{Res/Plast}$. This is consistent with the observations found in the literature (Fonseca *et al.*, 2016).

### 3.2 Influence of acoustic speckle on the nuclear magnetism relaxation times

Relaxation times $T_1$ and $T_2$ values evaluated at 1.5 T as a function of PVC and cellulose concentrations are reported in figure 2 (and further detailed in Appendix C). The presence of ultrasonic speckle does not appear to consistently affect $T_1$ and $T_2$ values of PVCP. $T_2$ values (averaged over all cellulose mass ratios) decrease by 10% (50 ms to 44 ms) between mass ratio between resin and plasticizer $M_{Res/Plast}$ from 40 to 50%; for $M_{Res/Plast}$ from 50% to 70%, no further change in $T_2$ is observed. The $T_1$ values of PVCP (averaged over all cellulose mass ratios) decrease from 258 ms to 223 ms with increasing mass ratio between resin and plasticizer $M_{Res/Plast}$ 40% to 70%. The results obtained at 1.5T are systematically compared to typical values found in the literature for similar PVCP TMP, most often at higher magnetic field, such as at 3T (He *et al.*, 2019) and at 7T (Li *et al.*, 2016). While $T_2$ values of PVCP are slightly shorter than those of healthy soft tissue at 1.5 T (54 ± 8 ms for liver, 35 ± 4 ms for muscle, 75 to 90 ms for brain, 90 ms for fat), $T_1$ values of PVCP are significantly shorter (600 ms for liver, 1060 ± 155 ms for muscle, 500 to 750 ms for brain, 200 ms for fat) (Allmann *et al.*, 1997; Graham *et al.*, 1999; Cieszanowski *et al.*, 2002; Gold *et al.*, 2004; Stanisz *et al.*, 2005; Kastler, B.; Vetter, 2011).

### 3.3 Mechanical properties of the PVCP phantom

The mechanical values obtained from indentation measurements up to 43 days after the beginning of the gelatin-fusion process are shown in figure 3(A). After a period of gelation-fusion of approximately 14 days, the linear mechanical properties are found to be stable for mass ratio between resin and plasticizer $M_{Res/Plast}$ 50% to 80%. These observations are consistent with the results obtained by MRE and dynamic mechanical analysis in (Arunachalam *et al.*, 2017). After a 2-week stabilization period, PVCP appears to be mechanically stable over time (for at least up to 4 weeks) at room temperature, without any specific storage conditions, such as immersion or moistening.

The mechanical values obtained from indentation measurements on day #1 of the gelatin-fusion process (purple) or averaged over all measurements obtained after day #14 once mechanical stabilization of the phantom is observed (orange) are reported in figure 3(B) and in Appendix C (for the immediate and long-term elasticity). The values are compared to typical values available from the literature for both brain (Bilston, Liu and Phan-Thien, 1997; Nicolle *et al.*, 2005; Hrapko *et al.*, 2006; Vappou *et al.*, 2007; Chatelin, Constantinesco and Willinger, 2010a) and liver tissues (Chatelin *et al.*, 2011). PVCP with a mass ratio $M_{Res/Plast}$ of 70% is an adequate candidate to mimic the elastic shear behavior of the liver tissue at small strain.

**Investigation of PVC plastisol tissue-mimicking phantoms for MR- and ultrasound-elastography**

In accordance with the mechanical time-evolution observed from indentation measurements, the ARFI and MRE acquisitions are performed directly after (1 day) and 43 days after beginning of the gelatin-fusion process. The same increase over time is observed as with indentation measurements. MRE results are presented in figure 4 and in Appendix C. The results are given in terms of mean shear modulus and their standard deviation in cylindrical regions of interest. The values are very close to those found by indentation.

The results from ARFI mechanical measurements are presented in figure 4 and in Appendix C and compared to indentation measurements. The values at day 1 are higher than those obtained by indentation (differences of 52%, 93%, 21% and 18% for $M_{Res/Plast}$ of 50%, 60%, 70% and 80%, respectively) and MRE (differences of 3%, 49%, 14% and 9% for $M_{Res/Plast}$ of 50%, 60%, 70% and 80%, respectively). The same trend is observed at day 43 with values from ARFI measurements higher than those obtained by indentation (differences of 83%, 23 %, 9 % and 4% for $M_{Res/Plast}$ of 50%, 60%, 70% and 80%, respectively) and MRE (differences of 32%, 22%, 12% and 2% for $M_{Res/Plast}$ of 50%, 60%, 70% and 80%, respectively). The higher the resin mass ratio is, the closest the results are between the different methods.

## 4  GENERAL DISCUSSION

The use of a single PVCP TMP for MRE and sonoelastography is investigated in this study. From our results, plastisol appears as a good candidate for mimicking soft tissues in terms of mechanical properties with $T_1$ and $T_2$ MR imaging values compatible with typical MRE pulse sequences. In addition, the ultrasound speckle can be adjusted by adding cellulose to PVCP without altering significantly its NMR properties. The current study provides indications for easily preparing PVCP phantoms for both MR and ultrasonography elastography, with readily available instruments, i.e. a microwave and a thermometer. Mechanical properties measured with indentation were found to stabilize 14 days after gelation-fusion, and remain mechanically stable until the end of our follow-up period of 43 days for PVCP gels with resin to plasticizer ratios of 50 to 80% and containing 0 to 1% mass ratio cellulose.

Many advantages for the use of PVCP in elastography phantoms can be listed, as attested by the current study: (1) short preparation time and ease-of-use; (2) fast solidification process, avoiding sedimentation of the acoustic speckle particles; (3) natural transparence to ultrasound, making it easy to control wave diffusion by adjunction of speckle particles; (4) long conservation time without any specific storage requirements, such as moistening or water bath required for other hydrogels such as PVA. Elasticity at small strains is simply controlled by varying the mass ratio between resin and plasticizer and consequently the PVC concentration. These specificities make it possible to combine both rheological and imaging properties close to those of biological soft tissues in a calibrated, robust TMP.

As previously reported, the acoustic attenuation of PVCP can mimic those of biological soft tissues through the addition of cellulose speckle particles. The results at 1MHz are in the typical range of biological organs (0.14 to 1.16 dB.cm$^{-1}$ at 1MHz (Maggi *et al.*, 2013)) and a cellulose concentration of 0.6% to 0.8% was found to mimic the acoustic attenuation of soft tissues. The mass ratio between resin and plasticizer $M_{Res/Plast}$ does not significantly influence the acoustic attenuation properties for concentrations higher than 50% in the studied PVCP. Limitations of this study in terms of measurement of the ultrasonic properties must be mentioned. First, the attenuation coefficient of PVCP TMP is characterized in this work at a single frequency (1 MHz), that lies below the center frequency of most clinical ultrasound transducers. How the studied PVCP TMP remains realistic in terms of acoustic

**Investigation of PVC plastisol tissue-mimicking phantoms for MR- and ultrasound-elastography**

attenuation over a wider frequency range remains unaddressed in this study. However, the values measured at 1 MHz agree with those found in the literature over a larger range of frequencies, as illustrated in figure 1. Second, the density and speed of sound used in this study were taken from the literature (Madsen *et al.*, 2003; Spirou *et al.*, 2005; Hungr *et al.*, 2012; Maggi *et al.*, 2013; Chakouch, 2016; De Carvalho *et al.*, 2016; Fonseca *et al.*, 2016; Vogt *et al.*, 2016; Dantuma, van Dommelen and Manohar, 2019; Bakaric *et al.*, 2020). They are here assumed to be the same for all the mass ratios of resin to plasticizer.

NMR relaxation times of the studied PVCP phantoms were found to slightly depend on the mass ratio between resin and plasticizer. None of the tested PVCP composition reproduces both the $T_1$ and $T_2$ at 1.5T of any biological soft tissues. The typical $T_2$ value of plastisol (45 ms) lies within the lower range of $T_2$ for biological soft tissues, including the liver and the heart. The $T_1$ values (220-260 ms) found for plastisol at 1.5 T are lower than those of most soft tissues. These relaxation times are different than those reported in the literature in PVCP without cellulose addition (Li *et al.*, 2016; He *et al.*, 2019), but direct comparison is not straightforward as measurement were performed at higher magnetic fields (3T and 7T, respectively) on different PVCP. However, the objective of this study was to evaluate whether the studied PVCP is a good TMP candidate for typical MRE acquisition sequences, rather than to reproduce NMR relaxation times of biological soft tissues. Since MRE relies on the use of phase images alone for the estimation of elasticity, and not on the $T_1/T_2$ weighted magnitude images, the studied PVCP can be used with typical MRE acquisition sequences, including the clinical hepatic MRE protocol, as demonstrated here.

PVCP gels are good surrogate for biological soft tissues from a mechanical point of view. Despite the same samples being tested across the different modalities, small differences in the stiffness values can be observed between MRE, ARFI and indentation measurements. These can be explained by the different physical protocols and assumptions between the three methods (harmonic shear wavelength, quasi-static compression and impulse shear wave velocity for MRE, indentation and ARFI, respectively). Consequently, the so-called "stiffness" does not exactly correspond to the same parameter under the same conditions. For instance, the frequency and strain rate ranges are intrinsic to the aforementioned methods and do not necessarily coincide. Previous investigations from the literature suggest that PVCP gels are likely to exhibit relatively low viscosity at the usual elastography frequency range and therefore moderate dependence of the elasticity with frequency (Hadj Henni *et al.*, 2011; Lefebvre *et al.*, 2016). Due to different experimental strain rate (or frequency), the viscosity could be a possible reason for the differences observed between indentation, ARFI and MRE measurements. It appears that the higher the resin concentration is, the closest the results are between the different methods. This could be explained by the fact that viscous effects are less predominant compared to pure elastic effects at higher concentrations, thus resulting in decreased shear wave dispersion and therefore less sensitivity to differences in mechanical frequency content. If needed, additional investigation should be carried out in order to evaluate whether including additional compounds could increase PVCP viscosity, as oil does in gelatin phantoms (Nguyen *et al.*, 2014).

The issue of cross-validation between elastography and rheological measurements or between different elastography methods is well known (Hamhaber *et al.*, 2003; Ringleb *et al.*, 2005; Vappou *et al.*, 2007; Green, Bilston and Sinkus, 2008; Oudry, Chen, *et al.*, 2009; Okamoto, Clayton and Bayly, 2011; Lefebvre *et al.*, 2016; Arunachalam *et al.*, 2017; Brinker *et al.*, 2018). As previously introduced, differences in frequency range and excitation modes –harmonic vs transient- make any direct comparison particularly difficult. PVCP with added cellulose offers the possibility of cross validating indentation with MR and US elastography methods. PVCP with added cellulose is a good, stable TMP

**Investigation of PVC plastisol tissue-mimicking phantoms for MR- and ultrasound-elastography**

for methods that require simultaneously both MR imaging and US attenuation for internally-generated displacements, such as MR-ARFI (Szabo and Wu, 2000; Souchon *et al.*, 2008; Vappou *et al.*, 2018).

The stability of the mechanical properties of a phantom over time is of great importance for the development, the validation and the comparison of elastography protocols. For instance, extending the shelf-life and the mechanical stability of gelatin and agar-agar hydrogels is challenging and requires adequate storage conditions and addition of preservative, fungicides or bactericide agents (Dang *et al.*, 2011). The plastisol TMP of this study were found to have stable mechanical properties, as measured with indentation testing, from day 14 until day 43 (end of follow-up period) after being manufactured. Over a longer follow-up period, the mechanical properties of a PVCP gel were found stable during storage, up to six months (Bohndiek *et al.*, 2013).

SEBS, PDMS and silicon exhibit similarly high shelf life with no specific storage constraints. Some silicone-based materials (such as the silicone-based Ecoflex gels (Hadj Henni et al., 2011)) share other similar properties and advantages with PVCP, and a similar characterization study of their mechanical and MR/US imaging properties would be of interest. Silicone and PDMS gels have relatively similar imaging, mechanical and storage characteristics as PVCP. Indeed, they all provide a solid support to which acoustic scatterers can be easily added, they are insoluble in water, remain stable during storage and have easily controllable properties (first of all their stiffness) (Fonseca *et al.*, 2016). While the sound velocity and acoustic attenuation values of PVCP are slightly lower and higher, respectively, than those of biological soft tissues, these limitations are even greater for silicone and PDMS, with values close to 1000 $m.s^{-1}$ (Culjat *et al.*, 2010) and over 9.8 $dB.cm^{-1}$ at 3 MHz (Tsou *et al.*, 2008), respectively. However, unlike silicone and PDMS gels, the preparation of PVCP gels requires curing at a given temperature range. In addition, access to the exact chemical composition of PVCP (PVC concentration, additives…) (Fonseca *et al.*, 2016) is manufacturer-dependent, which means that any direct comparison between different studies on PVCP should be performed with care. Another oil-based material, SEBS has been shown to mimic adequately acoustic (speed of sound between 1420 and 1464 $m.s^{-1}$ and acoustic attenuation between 0.4 and 4 $dB.cm^{-1}$ at 3.5 MHz, depending of the speckle particle density (Oudry, Bastard, *et al.*, 2009; Cabrelli *et al.*, 2017)) and mechanical properties of soft tissues with long term stability phantoms. As well as PVCP gels in this study, SEBS gels were shown to be as well efficient TMMs for both MRE and ultrasound elastography (Oudry, Chen, *et al.*, 2009). Although slightly lower than biological tissue values when compared to SEBS gels, we can however note that the use of PVCP TMM is characterized by the simplicity and speed of its preparation process.

## 5    CONCLUSIONS

This study provides indications about how to prepare a mechanically-stable phantom using plastisol in order to mimic biological soft tissues for MRE and/or sonoelastography. On the basis of their mechanical properties, PVCP phantoms are demonstrated to be realistic TMP with NMR relaxation times compatible to MRE. With the addition of cellulose particles, it is possible to develop a calibrated TMP for both MRE and sonoelastography measurements. Mechanical properties of the tested plastisol were found stable at room temperature after 2 weeks and at least until 6 weeks after gelation-fusion. For instance, a 60-70% mass ratio between resin and plasticizer $M_{Res/Plast}$ sample will mimic the linear elastic behavior of liver tissue correctly. In MRE, the $T_2$ values of the tested PVCP TMP are close to those of hepatic tissue, while with addition of 0.6% of cellulose particles, this gel becomes also a good liver-mimicking candidate in sonoelastography. Investigating the mechanical properties over a larger range of frequency using dynamic mechanical analysis (Lefebvre *et al.*, 2016; Arunachalam *et al.*, 2017) would be a great extension for this study.

# Investigation of PVC plastisol tissue-mimicking phantoms for MR- and ultrasound-elastography

## Conflict of Interest

The authors declare that the research was conducted in the absence of any commercial or financial relationships that could be construed as a potential conflict of interest.

## Author Contributions

SC, EB and JV wrote the manuscript. The PVCP TMP were prepared by SC. The indentation tests were developed and performed by SC, AA, LM, BW and JV. The acoustic test bench was developed, and the acoustic properties were investigated by SC, BW and JV. The NMR properties of the TMP were investigated by SC, EB and CG. SC performed final approval of the manuscript to be published and agreed to be accountable for all aspects of the work related to the accuracy and integrity. All authors agree to be accountable for the content of the work.

## Funding

This work has benefitted from support of the ANR (Agence Nationale de la Recherche) by the French national program "*Investissements d'Avenir*": (1) by the IDEX-Unistra (ANR-10-IDEX-0002-02) with the support of the IRIS technology platform; (2) by the LABEX-CAMI (ANR-11-LABX-004); (3) by the IHU Strasbourg (Institute of Image Guided Surgery, ANR-10-IAHU-02).

## Acknowledgments

The authors would like to thank the staff of the imaging platform of the IHU-Strasbourg for their support and availability and the IRIS technological platform of the ICube laboratory for technical support.

## Appendix A. Measurement of the acoustic attenuation coefficient

The acoustic attenuation coefficient is measured at 1 MHz for PVCP samples with varying mass ratio between cellulose and PVCP $M_{Res/Plast}$ using the set-up illustrated in the following figure:

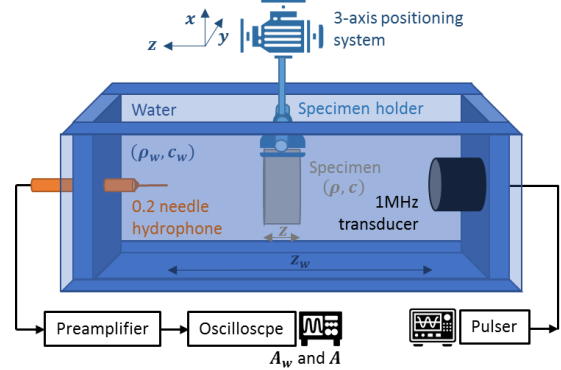

The acoustic attenuation of PVCP samples is investigated using the method proposed in (He and Zheng, 2001). The Fourier transform of a single-frequency plane wave propagating over two different media (water over a distance $z_w$ and a specimen over a depth $z$ - height of the sample) can be modeled as:

$$U(f) = A(f)e^{-i\theta} = U_0(f)e^{-(\alpha_w + i\beta_w)z_w}e^{-(\alpha + i\beta)z}T(f) \quad (1)$$

where $U_0(f)$, $f$, $\beta_w$, $\alpha_w$, $\beta$ and $\alpha$ are the Fourier transform of the initial generated pulse wave, its frequency, the propagation and the attenuation factors of water and the specimen, respectively.

# Investigation of PVC plastisol tissue-mimicking phantoms for MR- and ultrasound-elastography

$T$ is the overall water-specimen transmission factor defined by:

$$T(f) = \frac{4\rho c(f)\rho_w c_w}{(\rho c(f)+\rho_w c_w)^2} \quad (2)$$

with $\rho_w$, $c_w$, $\rho$ and $c$ density and speed of sound in water, the density of the specimen and the phase velocity, respectively. According to (He and Zheng, 2001), the attenuation coefficient is deduced from the amplitude spectra of the transmitted pulses with ($A$) or without ($A_w$) the specimen:

$$\alpha \cong \frac{\ln(T)}{z} + \frac{1}{z}\ln\left(\frac{A_w}{A}\right) \quad (3)$$

The amplitude drop is mostly expressed in dB by defining: $\alpha_{dB} = z\{20\log_{10}(e^{\alpha z})\} = 8.6886\alpha$. The attenuation coefficient of a material is generally dependent on the frequency $f$ of the ultrasound waves. A power law $\alpha_{dB} = af^\gamma$ can be assumed for this dependency (Duck, 1990; Szabo, 2004).

**Appendix B. Illustration of the indentation set-up.**

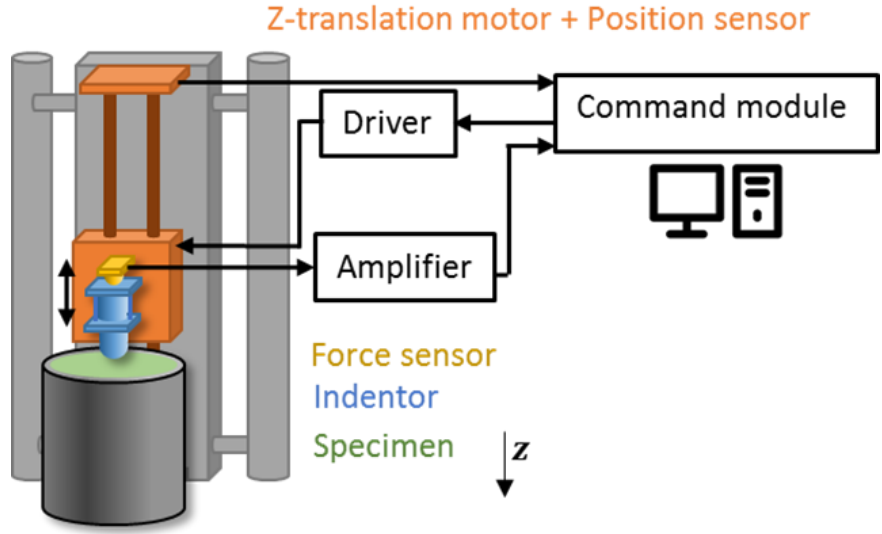

# Investigation of PVC plastisol tissue-mimicking phantoms for MR- and ultrasound-elastography

## Appendix C. Mechanical and imaging properties of PVCP

The main results obtained in this work on PVCP TMP are summarized in table A (mean and standard deviation values) for all tested properties (indentation, echographic and MRI properties).

**TABLE A.** Mechanical, ultrasonic and MRI properties of the tested PVCP

| | Mass ratio resin to plasticizer $M_{Res/Plast}$ [%] | | 40 | 50 | 60 | 70 | 80 |
|---|---|---|---|---|---|---|---|
| Echography | Mass ratio cellulose to PVCP $M_{Cell/Plast}$ [%] | | | | | | |
| | α [-] at 1MHz | 0.00 | 0.146 | 0.381 | 0.203 | 0.326 | - |
| | | 0.60 | 0.458 | 0.817 | 0.676 | 0.713 | - |
| | | 0.80 | 0.665 | 1.251 | 1.142 | 1.127 | - |
| | | 1.00 | 0.827 | 1.631 | 1.641 | 1.340 | - |
| | ARFI | $M_{Cell/Plast}$ [%] | | | | | |
| | Initial G [Pa] | 0.6 | 602 ± 109 | 772 ± 155 | 1901 ± 458 | 2913 ± 624 | 4529 ± 831 |
| | Long-term G [Pa] | 0.6 | 714 ± 221 | 1173 ± 150 | 2211 ± 288 | 3797 ± 636 | 5365 ± 593 |
| MRI | $T_1$ [ms] at 1.5T | $M_{Cell/Plast}$ [%] | | | | | |
| | | 0.00 | 263 | 253 | 234 | 223 | - |
| | | 0.60 | 257 | 247 | 233 | 222 | - |
| | | 0.80 | 255 | 247 | 235 | 223 | - |
| | | 1.00 | 258 | 255 | 241 | 224 | - |
| | | Mean $T_1$ [ms] | 258 ± 3 | 251 ± 4 | 236 ± 4 | 223 ± 1 | - |
| | $T_2$ [ms] at 1.5T | $M_{Cell/Plast}$ [%] | | | | | |
| | | 0.00 | 46 | 43 | 42 | 40 | - |
| | | 0.60 | 53 | 46 | 45 | 45 | - |
| | | 0.80 | 54 | 45 | 46 | 46 | - |
| | | 1.00 | 47 | 42 | 42 | 40 | - |
| | | Mean $T_2$ [ms] | 50 ± 3 | 44 ± 2 | 44 ± 2 | 43 ± 3 | - |
| | MRE at 60.1Hz | $M_{Cell/Plast}$ [%] | | | | | |
| | G [Pa] Day #1 | 0.6 | 272 ± 28 | 796 ± 113 | 1272 ± 120 | 2549 ± 438 | 4140 ± 266 |
| | G [Pa] Day #43 | 0.6 | 670 ± 84 | 890 ± 74 | 1810 ± 96 | 3401 ± 66 | 5240 ± 114 |
| Indentation | | $M_{Cell/Plast}$ [%] | | | | | |
| | G [Pa] Day #1 | 0.6 | - | 508 ± 109 | 985 ± 182 | 2410 ± 148 | 3836 ± 197 |
| | G [Pa] Day #43 | 0.6 | - | 640 ± 23 | 1803 ± 81 | 3480 ± 299 | 5155 ± 215 |

# Investigation of PVC plastisol tissue-mimicking phantoms for MR- and ultrasound-elastography

# Investigation of PVC plastisol tissue-mimicking phantoms for MR- and ultrasound-elastography

# Investigation of PVC plastisol tissue-mimicking phantoms for MR- and ultrasound-elastography

# Investigation of PVC plastisol tissue-mimicking phantoms for MR- and ultrasound-elastography

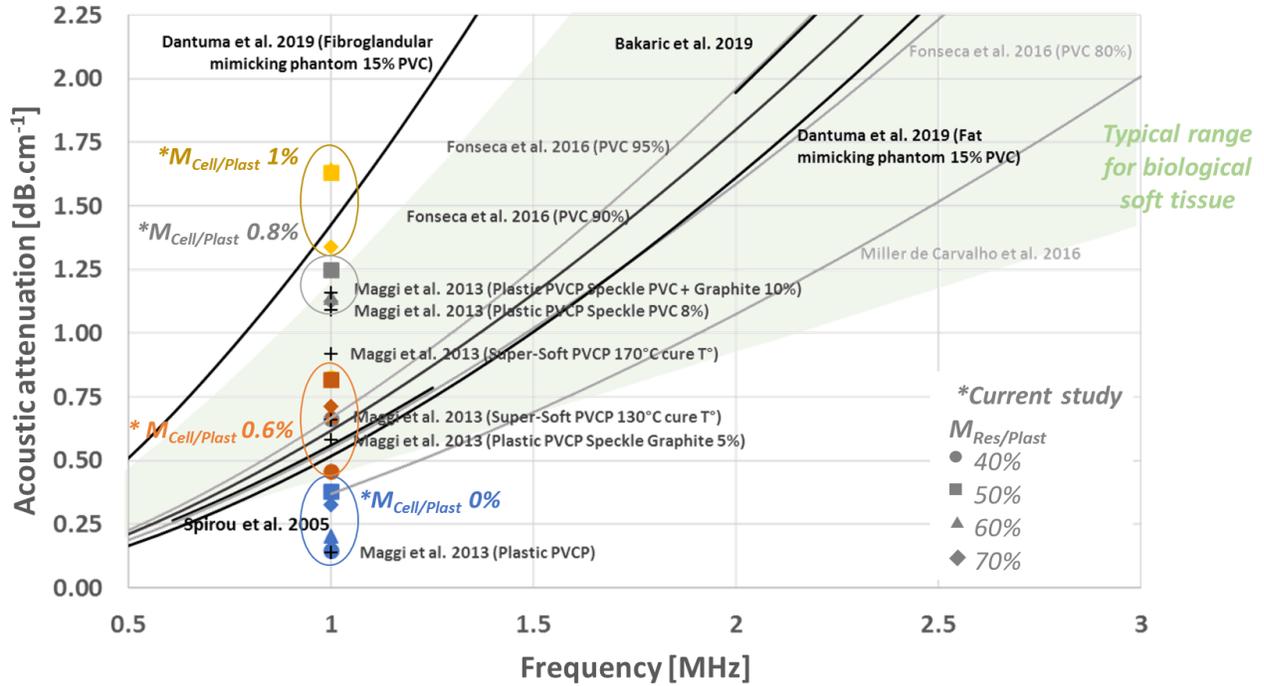

**figure 1.** The acoustic attenuation coefficients measured at 1 MHz in the tested Plastisol phantoms (colored signs) are compared to typical values from the literature for biological soft tissues (green colored area) (Duck, 1990) and to similar PVCP TMP (gray and black curves) (Spirou *et al.*, 2005; Maggi *et al.*, 2013; De Carvalho *et al.*, 2016; Fonseca *et al.*, 2016; Vogt *et al.*, 2016; Dantuma, van Dommelen and Manohar, 2019; Bakaric *et al.*, 2020) over the acoustic frequency range 0.5 to 3 MHz.

**Investigation of PVC plastisol tissue-mimicking phantoms for MR- and ultrasound-elastography**

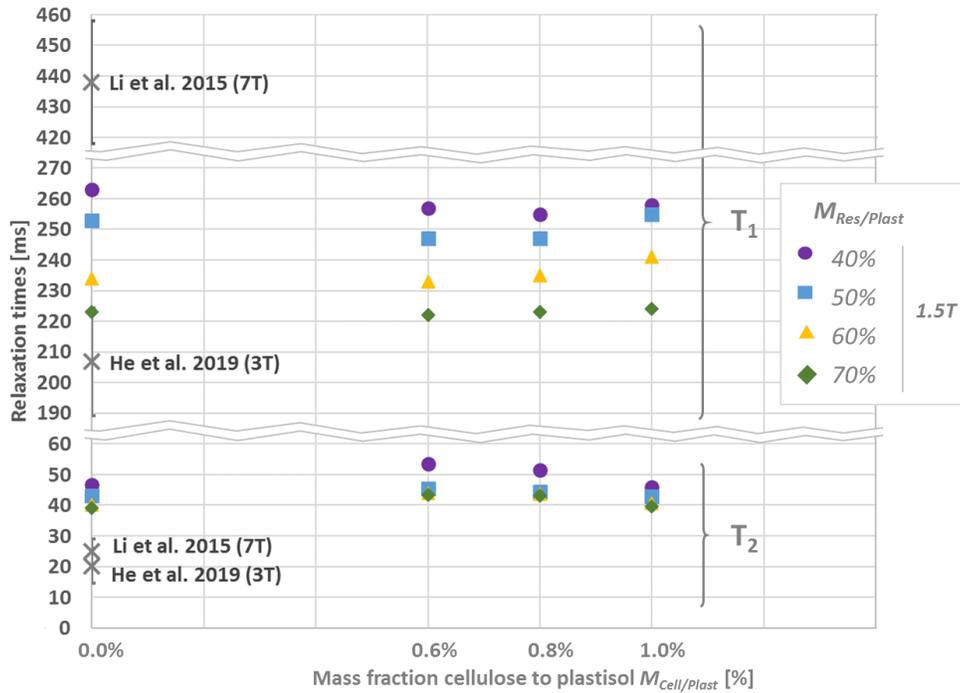

**figure 2.** Longitudinal $T_1$ and transverse $T_2$ NMR relaxation times acquired at 1.5T for varying mass ratio between cellulose and PVCP $M_{Res/Plast}$ and mass ratio between resin and plasticizer $M_{Res/Plast}$. These values are compared to $T_1$ and $T_2$ values from the literature measured in similar PVCP TMP (not containing cellulose) at 3T (He *et al.*, 2019) and 7T (Li *et al.*, 2016) (grey crosses).

**Investigation of PVC plastisol tissue-mimicking phantoms for MR- and ultrasound-elastography**

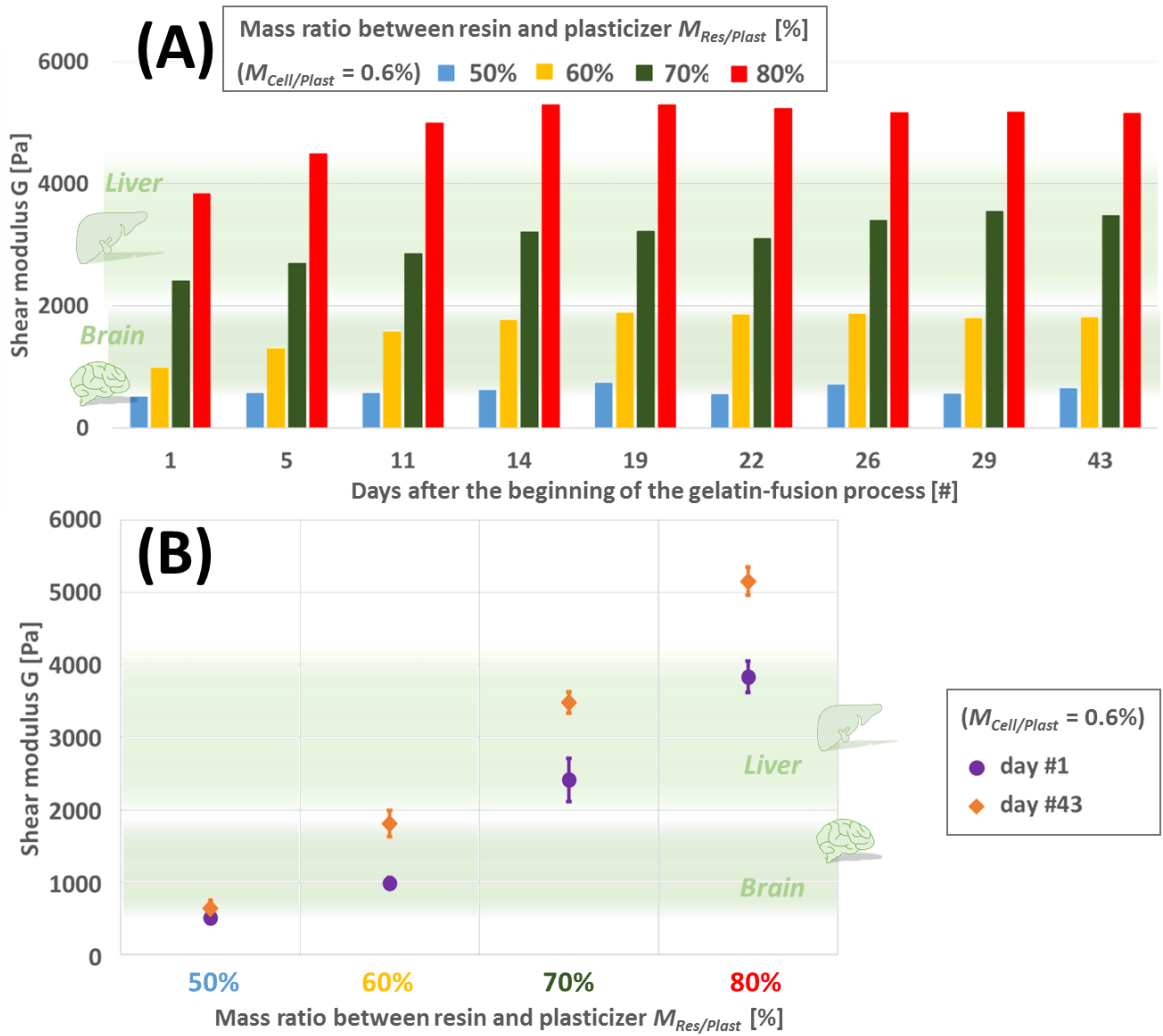

**figure 3.** The shear modulus against mass ratio resin to plasticizer is measured with indentation tests over a period of 43 days after gelation-fusion in order to study the stability of the PVCP gels in terms of mechanical properties (A). Shear moduli measured directly after gelation-fusion process (purple) and after stabilization (orange) are represented against mass ratio between resin and plasticizer, and are compared to liver (Chatelin *et al.*, 2011) and brain (Bilston, Liu and Phan-Thien, 1997; Nicolle *et al.*, 2005; Hrapko *et al.*, 2006; Vappou *et al.*, 2007; Chatelin, Constantinesco and Willinger, 2010b) values from the literature (B).

**Investigation of PVC plastisol tissue-mimicking phantoms for MR- and ultrasound-elastography**

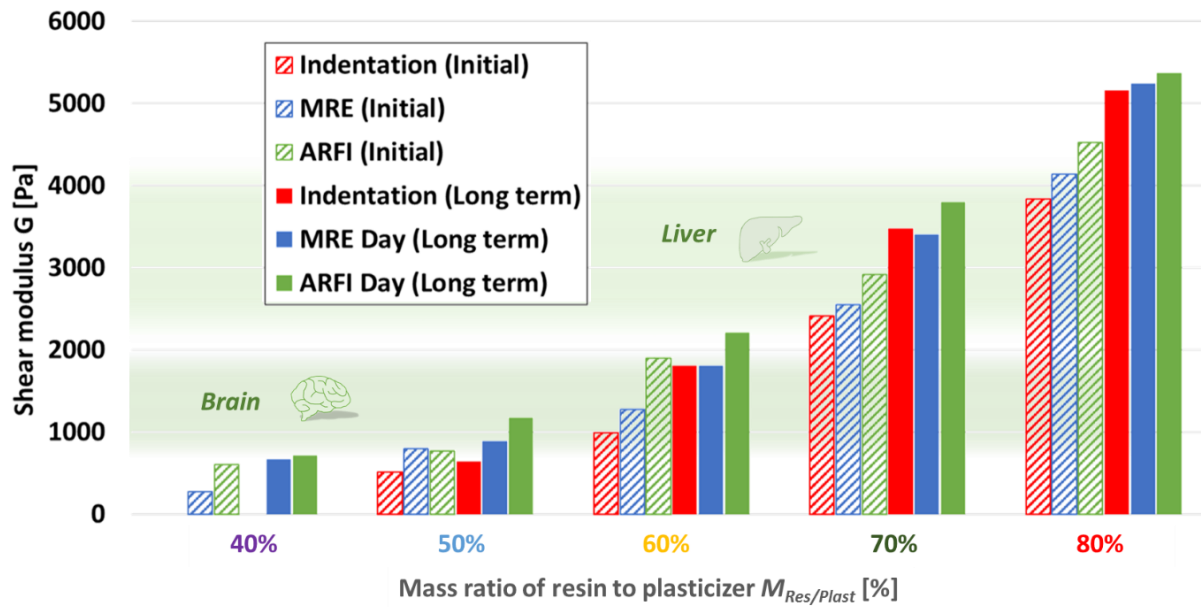

**figure 4.** Stiffness values measured directly ("Initial") and 43 days ("Long term") following the gelation-fusion process with MRE, ARFI and indentation testing. Stiffness ranges for liver (Chatelin et al., 2011) and brain (Bilston, Liu and Phan-Thien, 1997; Nicolle *et al.*, 2005; Hrapko *et al.*, 2006; Vappou *et al.*, 2007; Chatelin, Constantinesco and Willinger, 2010a) found in the literature are represented for comparison.